\documentclass[12pt]{iopart}

\def\half{{\textstyle{1\over 2}}}

\usepackage{bm}
\usepackage{graphicx}
\usepackage{multirow}
\usepackage{color}

\usepackage{iopams}
\begin{document}

\title[]{Discovery potential of hidden charm baryon resonances $ via$ photoproduction}

\author{Yin Huang$^{1,2,3}$, and Jun He$^{1,2,4}$\footnote{Corresponding
author: junhe@impcas.ac.cn},  Hong-Fei Zhang$^{3}$ and Xu-Rong Chen$^{1}$}

\address{
$^{1}$Institute of modern physics,Chinese Academy of Sciences, Lanzhou 730000, China\\
$^{2}$Research Center for Hadron and CSR Physics, Institute of Modern Physics of CAS
and Lanzhou University, Lanzhou 730000, China.\\
$^{3}$School of Nuclear Science and Technology, Lanzhou University,
Lanzhou 730000, China\\
$^{4}$State Key Laboratory of Theoretical Physics, Institute of
Theoretical Physics, Chinese Academy of Sciences}

\begin{abstract}
In this work, we study the possibility to find $N^*_{c\bar{c}}$ and
$\Lambda^*_{c\bar{c}}$ resonances with hidden charm with mass above 4
GeV  in the photon-induced production. The cross sections for the
photoproductions of hidden charmed baryons are predicted in the
effective Lagrangian approach with the vector meson dominance
mechanism. The $N^*(4412)$ can be produced directly by photon
excitation with total cross section about 1~nb. The $N^*(4412)$
provides a obvious peak near threshold for $J/\psi$ photoproduction,
which is promising to be checked by the future high precision
experiment at JLab 12~GeV. The results will be helpful to the
experimental search for the hidden-charmed baryon resonances in the
coming experiment at JLab 12~GeV, such as SoLID, and the proposed
electron-ion colliders at FAIR and HIAF.
\end{abstract}

\vspace{2pc}

\maketitle

\section{INTRODUCTION}

Searching for the exotic hadrons, which are the states beyond the
scheme of the conventional quark model with the figures $q\bar{q}$ and
$qqq$, is one of the main aims of the hadron physics. The most
promising candidates of the exotic states are the puzzled
charmonium-like states $XYZ$ observed in recent years, which are
difficult to be put into the frame of the conventional $c\bar{c}$
state~\cite{Godfrey:2008nc}. There exist various explanations about
such states, such as multiquark state, molecular states. In the baryon
side, historically the strange baryon $\Lambda(1405)$ was suggested to
be a molecular as $N\bar{K}$~\cite{Dalitz:1960du}. Recently, the
molecular state composed of charmed meson and nucleon is
discussed in the one-boson-exchange
model~\cite{Yasui:2009bz,He:2010zq,He:2012zd,Yamaguchi:2011xb}. The
hidden charm baryon resonances are also
suggested in the literature~\cite{Wu:2010jy,Wang:2011rga,Wu:2012md,Yang:2011wz,Yuan:2012wz,Gobbi:1992cf}.
In this work we
will focus on the hidden charmed baryon resonances.

In the conventional quark model, a baryon is made of three
constituent quarks. The lowest nucleon
resonance with spin-parity $J^P=1/2^{-}$ should be the first $L =1$
orbital excitation state while the experiment suggests the $N^{*}$ (1535) has higher mass than
the lowest $J^{P}=1/2^{+}$ radial excitation state $N^{*}$(1440).
Moreover, the experimental measurements about strange magnetic moment and
strange form factor of the proton also indict that the strange quark may
play an important role in the nucleon~\cite{Spayde:2003nr,Maas:2004dh}. Zou $et\ al.$ proposed to include
the pentaquark components as $qqqs\bar{s}$ in excited baryons to solve
such puzzles~\cite{Zou:2005xy}. It is natural to expect that there
exists baryons
with hidden charm if the excited energy becomes larger so that the
$c\bar{c}$ appears instead of $s\bar{s}$. Wu $et\ al.$ predicted the
narrow hidden charmmed bayon resonances $N^{*}_{\bar{c}c}$ and
$\Lambda^{*}_{\bar{c}c}$ with masses around 4.2$-$4.6~GeV and widths
smaller than 100 MeV, which are generated from $PB\to PB$ and $VB\to
VB$ coupled-channel systems with $P$ and $V$ the pseudoscalar and
vector mesons of the 16-plet of SU(4), respectively~\cite{Wu:2010jy}.
Since only S-wave is considered,  the angular momentum and parity
$J^P$ is set as $1/2^-$ in the original papers by Wu $et\
al.$~\cite{Wu:2010jy,Wu:2012md}.   Here we recall the predicted charmed
baryon resonances and their decay pattern in Table~\ref{Tab: NccWu}.
\begin{table}[h!]
\caption{The charmed
baryon resonances and their decay pattern predicted in
Ref.~\cite{Wu:2010jy} with unit of MeV. The first row is for the
generated channel. The third and fourth lines are for the masses $M$ and
total widths $\Gamma$, respectively. The $\Gamma_i$ is the partial
decay width of the corresponding  charmed
baryon resonance.  \label{Tab: NccWu}}
\renewcommand\tabcolsep{0.24cm}
\renewcommand{\arraystretch}{1.3}
\begin{indented}
\item[]
	\begin{tabular}{c|cr|crr|cr|crr}\hline
&\multicolumn{5}{c|}{$PB\to PB$}&
\multicolumn{5}{c}{$VB\to VB$}\\\hline
&&\multicolumn{1}{c|}{$N^*_{c\bar{c}}$}&
&\multicolumn{2}{c|}{$\Lambda^*_{c\bar{c}}$}&
&\multicolumn{1}{c|}{$N^*_{c\bar{c}}$}&
&\multicolumn{2}{c}{$\Lambda^*_{c\bar{c}}$}
\\\hline
$M$&& \multicolumn{1}{c|}{4261}&
& \multicolumn{1}{c}{4209}&
\multicolumn{1}{c|}{4394}  &
&\multicolumn{1}{c|}{4412}&
&\multicolumn{1}{c}{4368}  &
\multicolumn{1}{c}{4544}\\
$\Gamma$  &  &56.9 &     &32.4  &43.3  &     &47.3 &                  &28.0 &   36.6  \\\hline
&$\pi N$      &3.8  &   $K N$          &15.8  &0.0   &   $\rho N$      & 3.2 &   $K^* N$        &13.9 &   0.0     \\
&$\eta N$     &8.1  &   $\pi \Sigma$   &2.9   &10.6  &   $\omega N$    &10.4 &   $\rho \Sigma$  &3.1  &   8.8     \\
$\Gamma_i$&$\eta'N$     &3.9  &   $\eta \Lambda$ &3.2   &7.1   &   $K^*\Sigma$   &13.7 &   $\omega\Lambda$&0.3  &   9.1      \\
&$K\Sigma$    &17.0 &   $\eta'\Lambda$ &1.7   &3.3   &                 &     &   $\phi\Lambda$  &4.0  &   0.0               \\
&             &     &   $K\Xi$         &2.4   &5.8   &                 &     &   $K^{*}\Xi$     &1.8  &   5.0               \\
&$\eta_c N$   &23.4 &   $\eta_c\Lambda$&5.8   &16.3  &   $J/\psi N$    &19.2 &   $J/\psi\Lambda$&5.4  &   13.8  \\\hline
 \end{tabular}
 \end{indented}
\end{table}

It is interesting to search for such resonances in experiments with the help of the
partial decay width predicted. In the
original article by Wu $et\ al.$, the experimental search in the proton-antiproton
collision at PANDA has been suggested~\cite{Wu:2010jy}. On the other
side, it should be an ideal channel to produce the hidden charm
baryon resonances through the photon excitation by dragging out a pair of
$c\bar{c}$ from the nucleon. The photoproduction of the charmonium-like
states, such as $Z(4430)$ and $Y(3940)$ have been proposed and considerable
cross sections are predicted~\cite{Liu:2008qx,He:2009yda}.  There exist
many facilities in the world to study the normal nucleon resonances,
such as CLAS, MAMI, GRAAL and LEPS. However, all these facilities run
at too low energy to study hidden charm baryon resonances. The
proposed eRHIC, MEIC and LHeC run at too high energy. Fortunately, the
JLab 12 GeV with high luminosity just works in the energy region we
require. Besides, the proposed electron-ion colliders with lower
energy, such as ENC@FAIR in Germany and EIC@HIAF in China, can be used
to study the hidden charm baryon resonances. In Ref.~\cite{Wu:2012wta}, the generation of the bound state
$J/\psi-^3He$ is investigated in the photoproduction. In this work, we will
study the potential to find out such resonances in the photon-nucleon
interactions to give the helpful information for the future experimental
research.

The paper is organized as follows.  After the introduction, we will present
the formalism and ingredients used in our calculation. Then
in Section III, we will give the numerical results of the calculation, and the
possible background is also discussed.  The discussion and summary will be
given in the last section.

\section{Formalism for the photoproduction}

\begin{figure}[h!]
\begin{flushright}
\includegraphics[width=0.85\textwidth]{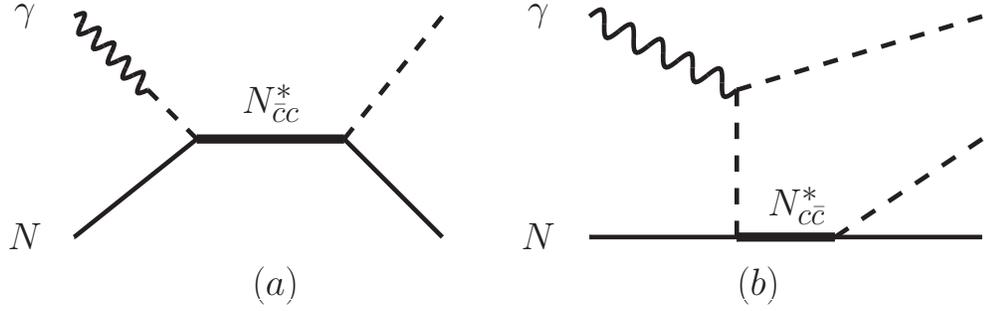}
\caption{The Feynman diagrams for the mechanisms of hidden baryon
	resonance production,
(a) direct production through VMD mechanism, (b) production through
$t$ channel. \label{Fig: diagram}}
\end{flushright}
\end{figure}

In this work we will study the photoproduction of the
hidden charm baryon $B^*_{{c}\bar{c}}$ including $N^*_{c\bar{c}}$ and
$\Lambda^*_{c\bar{c}}$, which can be divided into two types as shown
in Table~\ref{Tab: NccWu}: these decaying
decay to a vector meson and a baryon,  and these decaying to  a
peseudoscalar meson and a baryon. For the former the $N^*_{c\bar{c}}$
can be produced by vector meson dominance (VMD) mechanism shown in
Fig.~\ref{Fig: diagram} (a). The latter should be produced by a 
$t$ channel mechanism as shown in Fig.~\ref{Fig: diagram}
(b).

In the following, we will present the Lagrangians used in the
calculation. In the model used in Ref.~~\cite{Wu:2010jy}, only S-wave
is considered in the prediction of the mass and decay pattern of
hidden charmed baryon resonances. Here we adopt the Lorentz covariant
orbit-spin (L-S) scheme for the $B^*_{c\bar{c}}$
decays~\cite{Zou:2002yy}, with which only S-wave decay is involved in
the Lagrangian. In Ref.~\cite{Xie:2008ts}, Xie $et\ al.$ found the
L-S Lagrangian with VMD mechanism can give more consistent coupling constant of
$N^*(1535)N\rho$ with L-S coupling than the vector and tensor forms of coupling which
involve both S-wave and D-wave decays.  This scheme has also been used
in Refs.~\cite{Wu:2012md,Wu:2010rv} to study the production of
$B^*_{c\bar{c}}$ with $J^P=\half^-$ in electroproduction. The explicit  forms of the L-S Lagrangians used in
this work are
\begin{eqnarray}
{\cal L}_{B^*_{c\bar{c}}BP}&=&g_{B^*_{c\bar{c}}BP}~\bar{B}^*_{c\bar{c}}~P~B+h.c.,\\
{\cal L}_{B^*_{c\bar{c}}B V}&=&
ig_{B^*_{c\bar{c}}BV}~\bar{B}^*_{c\bar{c}}~\gamma_{5}
\gamma^\mu\tilde{g}^{\mu\nu}(p)~V^\nu~B+h.c.
\end{eqnarray}
where
$\tilde{g}^{\mu\nu}(p)=(g_{\mu\nu}-\frac{p_{\mu}p^\nu}{p^2})$ with $p$ the momentum of the charmed
baryon $B^*_{c\bar{c}}$. The explicit values of the couplings for hidden charmed baron
$B^*_{c\bar{c}}=(N^*_{c\bar{c}},\Lambda^*_{c\bar{c}})$, baryon $B=(N,\ \Sigma)$, pseudoscalar
meson $P=(\pi,\ \eta, \eta', K)$ and vector meson $V=(\rho,\ \omega,
J/\psi,\ K^*)$ can be found in
Table~\ref{Tab: ccNcc}. The
coupling constants can be determined by the partial decay widths
obtained in Ref.~\cite{Wu:2010jy} as listed in Table~\ref{Tab: NccWu}.
The obtained coupling constants which will be used in the following
calculation are presented in Table~\ref{Tab: ccNcc}.
\renewcommand\tabcolsep{0.42cm}
\renewcommand{\arraystretch}{1.3}
\begin{table}[h!]
\caption{The coupling constant for the decay of $N^*_{c\bar{c}}$ or
$\Lambda^*_{c\bar{c}}$.\label{Tab: ccNcc}}
\begin{indented}
\item[]\begin{tabular}{cc|cc|cc|cc}\hline
	\multicolumn{2}{c|}{$N^{*}(4261)$} & \multicolumn{2}{c|}{$N^{*}(4412)$}
	&	\multicolumn{2}{c|}{$\Lambda^{*}(4209)$} & \multicolumn{2}{c}{$\Lambda^{*}(4368)$}\\\hline	
$\pi N $       &0.103     &$\rho N$     &0.030 & $KN$ &0.369 & $K^*N$ & 0.125\\
 $\eta N$       &0.264     &$\omega N$   &0.096 &$--$&$--$&$--$&$--$\\
 $\eta^{'}N$   &0.189     &$J/\psi N$   &0.416 &$--$&$--$&$--$&$--$\\
 $K\Sigma$     &0.224     &$K^{*}\Sigma$&0.061 &$--$&$--$&$--$&$--$\\\hline
\end{tabular}
\end{indented}
\end{table}

To obtain the radiative decay vertex for $N^*_{c\bar{c}}$,
the VMD mechanism is introduced.
The $V\gamma$ coupling is described as,
\begin{eqnarray}
L_{V\gamma}=\sum_{V}\frac{eM_{V}^2}{f_{V}}V_{\mu}A^{\mu},
\end{eqnarray}
where $M_V$ is the mass of the vector meson. $V$ and $A$ are the vector meson
and electromagnetic fields. As suggested in Ref.~\cite{Xie:2008ts},
the gauge invariance of the diagram in Fig.~\ref{Fig: diagram} (a) can
be guaranteed by the propagator of the vector meson in VMD mechanism.

There are several ways to determine the coupling constants $e/f_V$.
In this work, we derive the coupling constants with the experimental partial decay width 
$\Gamma_{V\rightarrow{e^{+}e^{-}}}$ as listed in Table~\ref{Table: ccVMD}.
In the hidden-gauge approach~\cite{Nagahiro:2008cv,Branz:2010rj}, the coupling constants can be related to
tensor formalism~\cite{Ecker:1989yg}. In their
work, the $SU$(4)
symmetry-breaking effects are included by using
$g=M_{J/\psi}/(2f_{\eta_c})$ for $\gamma-J/\psi$ coupling. The values
are also presented in Table~\ref{Table: ccVMD}.  
\renewcommand\tabcolsep{0.2cm}
\renewcommand{\arraystretch}{1.3}
\begin{table}[h!]
\caption{The coupling constant $e/f_{V}$ determined in $V\rightarrow{}e^{+}e^{-}$,
The data for branching rations are from PDG~\cite{PDG}. The values in
the third column are from the hidden-gauge approach~\cite{Branz:2010rj}. \label{Table: ccVMD}}
\begin{indented}
\item[]\begin{tabular}{ccccc}\hline
		Coupling constant     & Values             &
	Ref.~\cite{Branz:2010rj}  &Total width   &$BR(V\rightarrow{e^{+}e^{-}})$\\
$e/f_{V}$             &$(\times{}10^{-2})$ & $(\times{}10^{-2})$  & (MeV)        &$\times
10^{-5} $\\
\hline
$e/f_{\rho}$        &$4.986$  & $5.173$   &$149.1\pm0.8$
&$4.72\pm0.05$\\
$e/f_{\omega}$      &$1.459$  & $1.698$   &$8.49\pm0.08$
&$7.28\pm0.14$\\
$e/f_{J/\psi}$      &$2.209$  &$3.873$    &$(92.9\pm2.8)\times10^{-3}$
&$(5.94\pm{}0.06)\times{}10^{3}$\\\hline
\end{tabular}
\end{indented}
\end{table}

For the diagram (b) in Fig~\ref{Fig: diagram}, a
typical effective Lagrangians for the $\gamma PP$ and $V\gamma P$
couplings are introduced in the following form,
\begin{eqnarray}
&L_{\gamma{}\pi{}\pi}=-ie[\pi{}\partial_{\mu}\pi^{+}-\pi^{+}\partial_{\mu}\pi]A_{\mu},\\
&L_{V\gamma{}P}=g_{V\gamma{}P}\epsilon_{\mu\nu\alpha\beta}\partial^{\mu}V^{\nu}\partial^{\alpha}A^{\beta}P
\end{eqnarray}
where $A$, $V$ and $P$  are electromagnetic, vector meson and
pseudoscalar fields, $M_{V}$ is the vector meson mass,
$\epsilon_{\mu\nu\alpha\beta}$ is the anti-symmetric Levi-Civita
tensor. The coupling constants $g_{K^{*+} K^+\gamma}=0.254$~GeV$^{-1}$
and $g_{\rho^+\pi^+\gamma}=0.222$~GeV$^{-1}$ are extracted from the
experimental decay width~\cite{PDG}.

To include the off-shell effect, two types of form factors
for meson and baryon in the $BBM$ vertices are introduced,
\begin{eqnarray}
	F_{M}(q^2)=\frac{\Lambda_{M}^{2}-m_{M}^2}{\Lambda_{M}^{2}-q^2},\quad
	F_{B}(q^2)=\frac{\Lambda_B^4}{\Lambda^4+(q^2-m_B^2)^2},
\end{eqnarray}
where the $m_{(M,B)}$ is the mass of the off-shell meson or baryon.
In this work the cut off $\Lambda_{M}$ is chosen as
$\Lambda_{M}=m_M+\alpha \Lambda_{\rm QCD}$ with
$\Lambda_{\rm QCD}=0.22$~GeV. The parameter $\alpha$ is close to
unitary usually. In this work, we choose $\alpha=1.3$. The cutoff for the off-shell baryon is
chosen as $\Lambda_B=1$~GeV. In this work we will also discuss the
effect of the uncertainties of the cut off by variation $\alpha$ from
to 2.

\section{Numerical results}

In this section we will study the production of the hidden charmed
baryon with the Lagrangians given in the previous section  with the
possible background from the experiment or theoretical prediction.

\subsection{The photoproduction of $N^{*}(4412)$}

Since $N^*(4412)$ decays to $N/\Sigma$ and a vector meson
mainly, the photon can excite the target nucleon to
the hidden charmed nucleon resonance $N^{*}(4412)$ through VMD mechanism  as shown
in Fig~\ref{Fig: diagram} (a). The intermediated vector mesons in the
VMD mechanism are the neutral vector meson $\rho^0$, $\omega$ and $J/\psi$. Due to the off-shellness of the
intermediate vector mesons, the form factor should be added to the
corresponding vertices~\cite{Zhao:2006gw}, which will
lead to the suppression of  $J/\psi$ exchange contribution.

As shown in Table~\ref{Tab: NccWu}, $N^*(4412)$ only couples
to $VB$, which is due to the exclusion of the $VB\to PB$ transition in
Ref.~\cite{Wu:2010jy} The $VB\to PB$ transition would involve
pseudoscalar exchange. However, the studies in the light sector for
baryon resonances~\cite{Khemchandani:2011mf,Garzon:2012np} and in the
heavy sector for hidden charm mesons have been done concluding that
this effect would be small~\cite{Molina:2009ct,Nieves:2012tt}. Hence
we do not consider this correction in this work.  Other possibility is
a contact term such as the Kroll-Ruddermann term studied in
Refs.~\cite{Khemchandani:2011mf,Garzon:2012np}. It should a background
contribution which will be included in two-gluon and three gluon exchange
contributions.

\begin{figure}[h!]
\begin{flushright}
\includegraphics[width=0.8\textwidth]{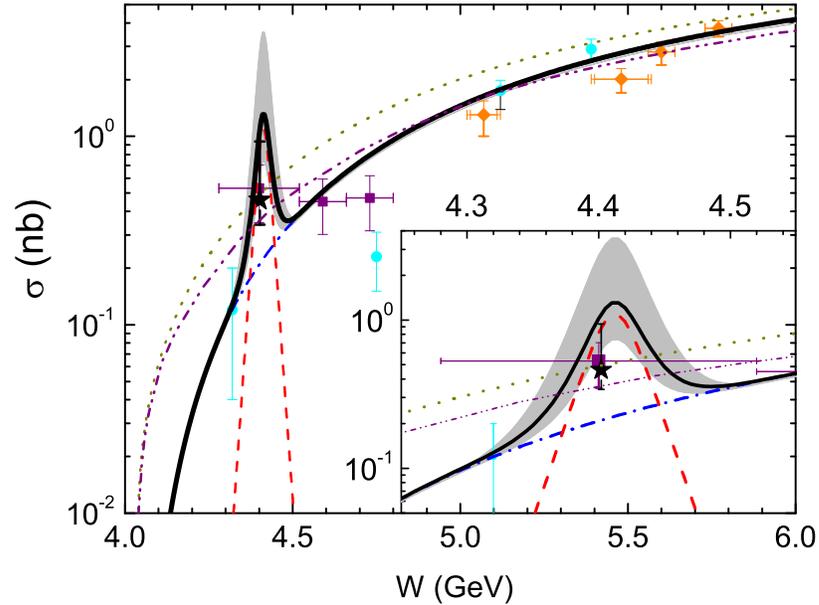}
\caption{(Color online) The cross sections for $J/\psi$ photoproduction	as function of the energy $W$ in center of mass frame
	.The full (black), dashed (red)
dash-dotted (blue) are for total, $N^*(4412)$ and
	back ground contributions. The band is for the total cross section with
	the variation of $\alpha$ in the cut off $\Lambda$ from 1 to
	2. The star is explained in the text. The $2g+3g$ model and PM
	model results in Ref.~\cite{Wu:2012wta} are also shown as
	dotted (dark yellow) and dash-dot-dotted (purple) lines.The subfigure focus on
	the energies around $W=$4.412 GeV, and
	the position of the star is moved to the right slightly to avoid
	overlapping with the experimental data from Cornell. The experimental data are from SLAC
	75~\cite{Camerini:1975cy} (square), SLAC 76~\cite{SLAC76}
	(diamond)  and 	Cornell~\cite{Gittelman:1975ix} (circle). \label{Fig: Ncc4412tcs}}
\end{flushright}
\end{figure}

The near threshold behavior of $\gamma p \to J/\psi p$ process has been studied in the
literature~\cite{Brodsky:2000zc,Sibirtsev:2004ca,Wu:2012wta}. Here we adopt the
two gluon (2$g$) and three-gluon (3$g$) exchanges in Ref.~\cite{Brodsky:2000zc} as
background.  In Ref.~\cite{Wu:2012wta}, the $2g+3g$ model can describe
the cross section up to an energy about 20 GeV generally. The
amplitude adopted is written as~\cite{Wu:2012wta},
\begin{eqnarray}
	{\cal
	A}&=&\frac{1}{2\pi^2}\frac{1}{2E_{J/\psi}(|{\bm
	k}|)}\sqrt{\frac{m_p}{E_p(|{\bm k}|)}}
	\sqrt{\frac{m_p}{E_p(|{\bm q}|)}}\frac{1}{\sqrt{2|{\bm q}|}}
	\frac{4\sqrt{\pi}}{\sqrt{6}}\frac{|{\bm q}|W}{m_p}[{\bm M}_{2g}+{\bm M}_{3g}]
\end{eqnarray}
with
\begin{eqnarray}
	{\bm
	M}_{2g}=\frac{A_{2g}}{4\sqrt{\pi}}\frac{1-x}{RM_{J/\psi}}
	e^{bt/2},\quad
	{\bm
	M}_{3g}=\frac{A_{3g}}{4\sqrt{\pi}}\frac{1}{R^2M^2_{J/\psi}}
	e^{bt/2},\label{Eq: 23g}
\end{eqnarray}
where $E_{J/\psi}(|{\bm k}|)$, $E_p(|{\bm k}|)$, $E_p(|{\bm q}|)$ are the energies for the
$J/\psi$, initial proton and final proton with initial and final
momentum $k$ and
$q$. $x=(2m_pM_{J/\psi}+M^2_{J/\psi})/(W^2-m_p^2)$, where $W$,  $M_{J/\psi}$ and $m_p$ are the energy in center of mass
frame, the mass of $J/\psi$ and mass of proton.
As in Refs.~\cite{Brodsky:2000zc,Wu:2012wta}, $R=1$~fm and
$b$=1.14~GeV$^{-2}$. The parameters $A_{2g}$ and $A_{3g}$ are
determined by fitting the data. With the background and resonance
contributions, the cross section can be obtained as shown in
Fig.~\ref{Fig: Ncc4412tcs}.

The old experimental data with large uncertainties, which were obtained
in the seventies of the last century, show a peak near the energy point
$W=4.412$~GeV.  Althrough the data point about 4.4~GeV is well
reproduced in the $2g+3g$ model in Ref.~\cite{Wu:2012wta},
the cross sections in the $2g+3g$ model are larger than the PM model and
experimental data generally.  If we decrease the $3g$ contribution, the
experimental data can be well reproduced with almost only the $2g$
exchange mechanism except the data point at $W=4.4$ GeV.  In
Ref.~\cite{Brodsky:2000zc}, Brodsky $et\ al.$ suggested that inclusion
of the contributions of five-quark resonance near threshold will
improve the theoretical results. Here we consider the hidden charm
baryon resonance $N^*(4412)$, which gives a narrow peak at energies
about 4.4 GeV as shown in Fig.~\ref{Fig: Ncc4412tcs}. To fit the
experimental data, the $3g$ contribution should be suppressed by
adjusting the free parameter $A_{3g}$. After including the
$N^*(4412)$, the $2g$ contribution is dominant to reproduce the
experimental data. Hence we consider the $2g$ contribution only  in
this work and determine the parameter $A_{2g}$ with the experimental
data except the data point around 4.4 GeV. It is still possible that
the $3g$ contribution is dominant very near threshold (such as lower
than 4.2 GeV), which is out of the interesting energy scope in the
current work. In Ref.~\cite{Wu:2012wta}, the
PM model is fitted to all data points around 4.4 GeV without
explicit mechanism at energies around 4.4 GeV. Hence, though there are
some differences at low energies, our
result is not in conflict with the PM model.

The contribution from the charmed baryon $N^*(4412)$ can be calculated
from the predicted decay width in Ref.~\cite{Wu:2010jy} and
Lagrangians presented in the previous section. Due to the narrow width
of $N^*(4412)$, we give the total cross section in the corresponding
bin of the Cornell experiment to make comparison. The result shows
that
after including the contribution of the $N^*(4412)$ to the background
contribution, the theoretical result with $\alpha=1.3$ is close to the
experimental one. However the large uncertainty of the existing data
make it difficult to give a confirmative conclusion. The results are not sensitive
to the cut off $\Lambda_B$. To present the effect of the variation of
the parameters $\Lambda_M=m_M+\alpha \Lambda_{QCD}$, the results with
$1<\alpha<2$ are also shown in the Fig.~\ref{Fig: Ncc4412tcs}. The
theoretical results are consistent with the experimental data
considering the uncertainty of the experiment.  The confirmation of
the existence of $N^*(4412)$ need high precision experiment. The total
cross section through intermediating $N^*(4412)$ is about 1~nb at the
energy point $W=4.412$~GeV.  An experiment with precision about 0.1~nb
will be enough to check the existence of the hidden-charmed baryon
$N^*(4412)$.  The future experiment in JLab 12~GeV with large
luminosity is promising to reach such requirement~\cite{ATHENNA}.

\begin{figure}[h!]
\begin{flushright}
\includegraphics[width=0.85\textwidth]{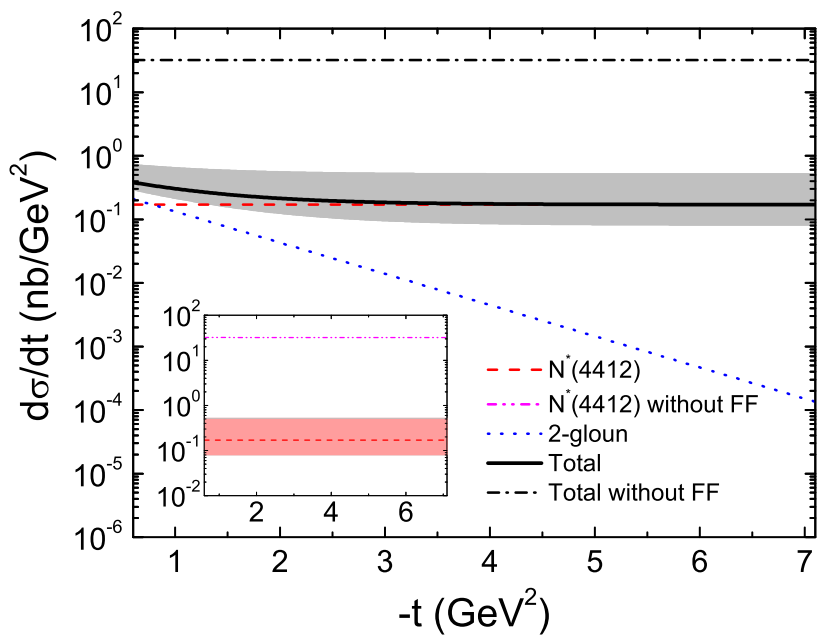}
\caption{
(Color online) The differential cross sections for $J/\psi$ production as function of $-t$ in center of
mass frame. The full, dashed and dotted lines are for the total,
$N^*(4412)$ and $2g$ contributions, respectively. The grey band is
for total contribution with the variation of $\alpha$ from 1 to 2.  The subfigure is
for the $N^*(4412)$ contribution with $1<\alpha<2$. The dash-dotted
and dash-dot-dotted lines are for the total and $N^*(4412)$
contributions without from factor (FF).
\label{Fig: Ncc4412dcs} }
\end{flushright}
\end{figure}

In the former results the total cross sections are given, now we present the
differential cross sections at the energy point $W=4.412$~GeV in
Fig.~\ref{Fig: Ncc4412dcs}. It is well known that the differential
cross section $d\sigma/dt$ will decrease exponentially as suggested
by Eq.~(\ref{Eq: 23g}). The existence of the $N^*(4412)$ will
change the feature of the differential cross section completely. As
shown in Fig.~\ref{Fig: Ncc4412dcs}, the differential cross section
$d\sigma/dt$ is weakly dependent of Mandelstam variable $-t$. Since $-t$ is in proportional to the
outgoing angle $\theta$ at certain $W$, the differential cross
section $d\sigma/d\cos\theta$ is also weakly dependent of angle $\theta$  in the center of mass frame.
Hence, it is an excellent way to search for the charm baryon resonance
$N^*(4412)$ by measuring the differential cross sections at the energy
point $W=4.412$~GeV with higher $-t$.

We also present the results without the form factors to show the
uncertainties tied to the use of the form factors. The results show
that the form factor can lead to the suppression of the cross section
by two orders of magnitude. If the form factors are removed,
there are large deviations from the experimental data even if the
large uncertainties of the data are considered. Hence, the inclusion
of the form factors is necessary in our effective Lagrangian
approach.

Besides the $J/\psi p$ channel, we will consider the $K\Sigma$ channel.
In the  $K\Sigma$ channel there are no experimental data in the energy
region near $W=4.412$~GeV to our knowledge. There exist many
theoretical studies~\cite{Ader:1970gi,Guidal:1997hy,Sibirtsev:2009bj}. To estimate the background
, we extend the theoretical results for the $\gamma p\to
K^{*0}\Sigma^+$ and $\gamma p\to K^{*+}\Sigma^0$ by the model of Nam
$et\ al.$~\cite{Kim:2012pz} to the higher energy region as the background
for the hidden charm $N^*(4412)$. At the low energies considered in
Ref.~\cite{Kim:2012pz}, the dominant contributions are from $K$-,
$\kappa$-exchanges and $\Delta$. At energies higher than 4~GeV,
the contributions from s-channel, that is, the contributions from the
normal nucleon resonances, should be negligible because their masses are
below 3~GeV. The t-channel contributions from $K^*$, $K$ and $\kappa$
exchanges are dominant and the contribution from  $\Delta$ is
negligible. For the neutral $K^{*}$ photoproduction, the contact and
$K^*$ exchange contributions are absent. For the charged $K^*$
photoproduction, the contact contribution decreases slowly and can
provide considerable contribution as shown in  Fig.~\ref{Fig:
Ncc4412KsS}.

\begin{figure}[h!]
\begin{flushright}
\includegraphics[width=0.85\textwidth]{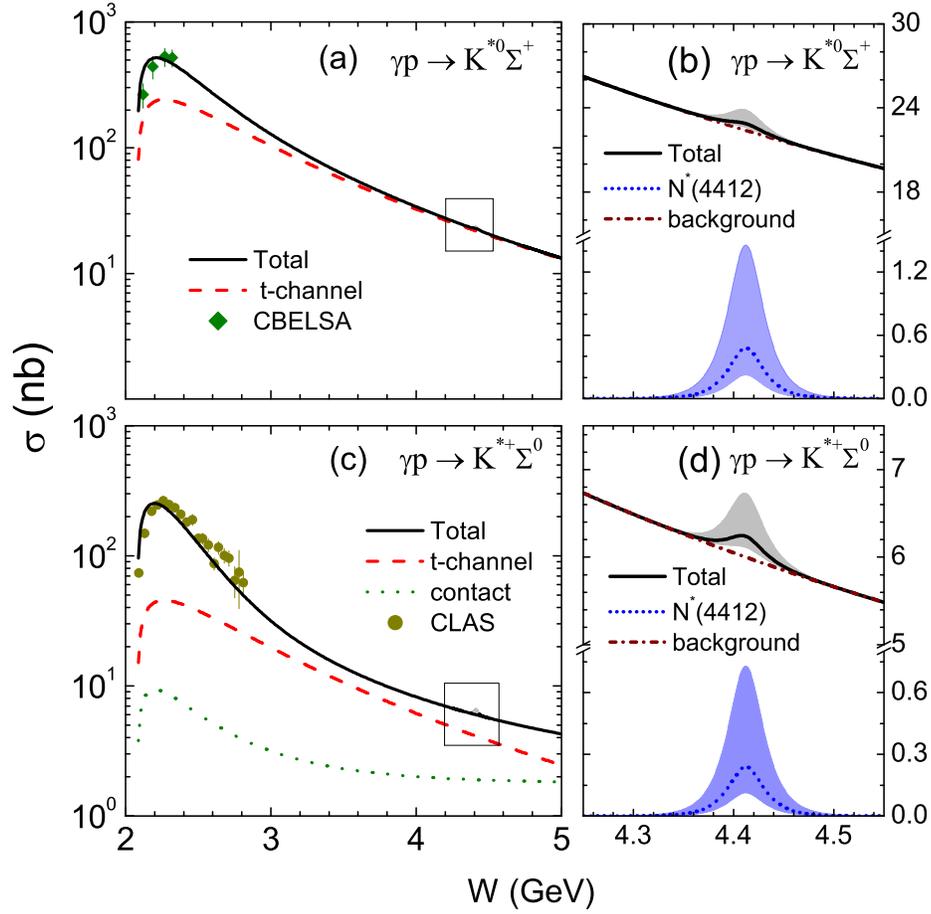}
\caption{
(Color online) The total cross sections for $\gamma p \to K^{*+}\Sigma^0$ and $\gamma
p\to K^{*0}\Sigma^+$ as function of the energy $W$ in the center of mass
frame. The figures (a) and (c) are for the cross sections at the
energies from threshold to 5 GeV of proceeses $\gamma p \to K^{*+}\Sigma^0$ and $\gamma
p\to K^{*0}\Sigma^+$, respectively.  The figures (b) and (d) focus on
the energies near the $N^*(4412)$ pole. And the contribution of
$N^*(4412)$ is presented in figures (c) and (d) as a dotted line.  The grey and light blue bands are for total and
$N^*(4412)$ contributions.  The experimental data are from
Refs.~\cite{Tang:2013gsa,Nanova:2008kr}.
	\label{Fig: Ncc4412KsS} }
\end{flushright}
\end{figure}

The background contributions at the energy point
$W=4.412$~GeV are about 20~nb and 6~nb for the $\gamma p\to
K^{*0}\Sigma^+$ and $\gamma p\to K^{*+}\Sigma^0$ processes,
respectively. As shown in Fig.~\ref{Fig: Ncc4412KsS}, the
total cross sections from $N^*(4412)$ at the energy point
$W=4.412$~GeV with $\alpha=1.3$ are $0.4$~nb and $0.2$~nb for
$K^{*0}\Sigma^+$ channel and $K^{*+}\Sigma^0$ channel, respectively.
With the variation of the cut off, the total cross section from the
$N^*(4412)$ is in the
order 0.1-1 nb.
Compared with the $J/\psi p$ channel, the total cross
section from  $N^*(4412)$ is smaller while the background is larger.
The precision required in this channel is of the order of 0.01~nb.
Hence the requirement to observe $N^*(4412)$ in this channel is higher
than that in $J/\psi p$ channel.

The total cross sections of $N^*(4412)$ photoproductions in the $\rho p$
and $\omega$ channels can be calculated analogously. In the $\rho p$
channel the total cross section is about one order of magnitude smaller than the
$J/\psi p$ and $K\Sigma$ channels, which results in that the accuracy
required in the search for $N^*(4412)$ in the $\rho p$ channel is very
high. The total cross section for the $\omega p$ channel is of the same order
as the channels $J/\psi p$ and for $K\Sigma$. There exists some old experimental data in
the energy region considered in this work. The SLAC
experiment done in the seventies of last century measured the total cross
section at the photon energy $E_\gamma=9.3$~GeV, that is, the energy
in center of mass frame  $W=4.3$~GeV,  and values about 20 $\mu$b and
5~$\mu$b are obtained~\cite{Ballam:1972eq}. Such large background make
the search for $N^*(4412)$ very difficult
considering the small produced cross section predicted in the current
work. Hence we do not suggest to search $N^{*}(4412)$ in $\gamma p\to
\rho p$ and $\gamma p\to \omega p$ channels.

\subsection{The photoproduction of $N^{*}(4261)$}

\begin{figure}[h!]
\begin{flushright}
\includegraphics[width=0.85\textwidth]{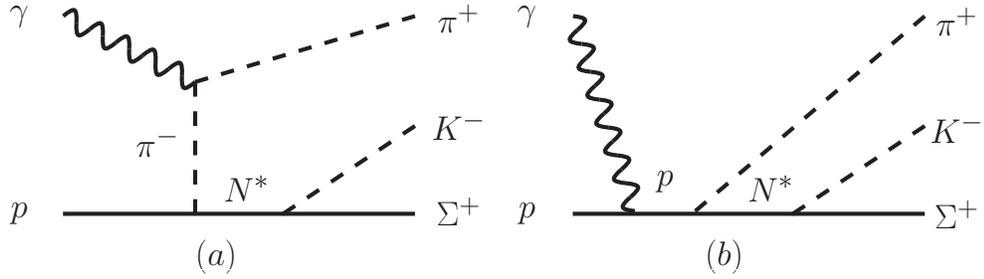}
\caption{The diagram for $\gamma p\to \pi^+ K^{-}\Sigma^+$. Figures (a) and
(b) are for t-channel and s-channel, respectively. \label{Fig: N4261}}
\end{flushright}
\end{figure}

Now we turn to another predicted hidden charm nucleon resonance,
$N^{*}(4261)$, which decays to a pseudoscalar meson and a baryon
mainly. Due to that a vector meson is essential in the VMD mechanism,
$N^{*}(4261)$ can not be produced by photon excitation directly. It
should be produced through the mechanism shown in Fig~\ref{Fig:
diagram}(b). Considering that the $\pi$ meson is easy to be detected in the
experiment we consider the final state with $\pi$ and $N^*(4261)\to
BP$ only. Since the vertex $\gamma \pi^0\pi^0$ does not exist, the
only possible channel is $\gamma p\to \pi^+N^{*}_{c\bar{c}}\to \pi^+
BP$ as shown in Fig.~\ref{Fig: N4261} (a). However the amplitude with
only $t$ channel contribution does not satisfy the gauge invariance. The
$s$ channel contribution as shown in Fig.~\ref{Fig: N4261} (b) is
included to restore the gauge invariance. Compared with the
$N^*(4412)$ which can be produced by the photon
excitation directly through VMD mechanism, the possibility for the production of
$N^*(4261)$ with three body final state should be smaller due to the
suppression of the phase space.

\begin{figure}[h!]
\begin{flushright}
\includegraphics[width=0.8\textwidth]{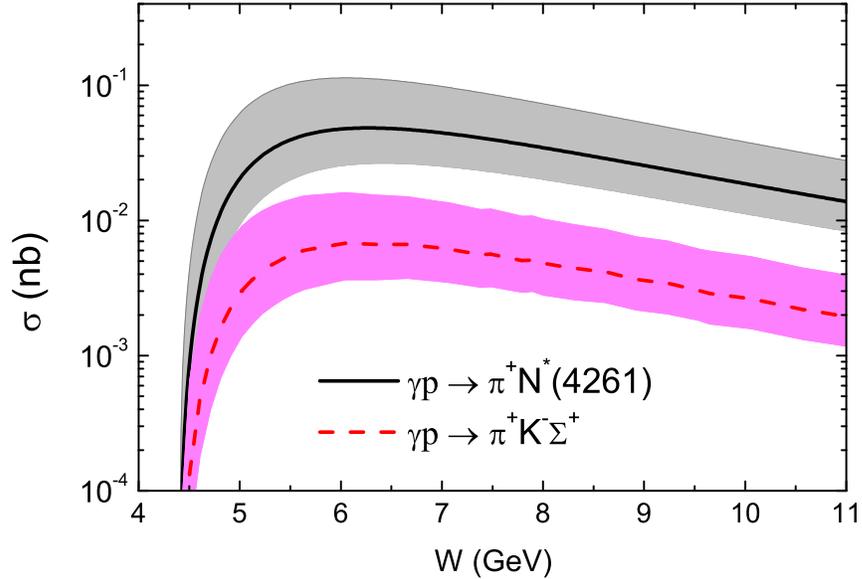}
\caption{(Color online) The total cross sections for photoproduction
	of $N^*(4261)$ as a function of the energy $W$ in center of mass frame
	. The full and dashed lines are for the
	total cross section of $\gamma p\to \pi^+ N^*(4261)$
	and $\gamma p\to \pi^+
	K^-\Sigma^+$ channel. The bands are for the uncertainties
with the variation of $\alpha$ from 1 to 2. \label{Fig: Ncc4261dcs} }
\end{flushright}
\end{figure}

First, we consider the total cross section for the $N^*( 4261)$
production, $\gamma p\to \pi^+ N^*(4261)$. As shown in Fig.~\ref{Fig:
Ncc4261dcs}, the total cross sections are in the order 0.01-0.1~nb
which are about one oder of magnitude lower than the total cross sections of
$N^*(4412)$.  The total cross sections for the channels $\pi^+
K\Sigma$ and $\pi^+\eta^{(')}_{(c)} N$ can be estimated from the total
cross section for the $N^*(4261)$ production with the help of the
branch ratio of the corresponding decay channel. Here we take
the channel $\gamma p\to \pi^+ K^-\Sigma^+$ as an example to make an
explicit calculation.In Fig.~\ref{Fig: Ncc4261dcs}, total cross
section from $N^*(4261)$ is of the order of 0.001 nb. In the
literature~\cite{Erbe:1970cq,Erbe:1967,CambridgeBubbleChamberGroup:1967zzb},
there existed some old data for  $\gamma
p\to \pi K\Sigma$, which are at the order of $\mu b$.  It is
difficult to search for the  $N^*(4261)$ in the experiment with so large
background.

\subsection{The photoproduction of $\Lambda^{*}_{\bar{c}c}$}

The hidden charm strange baryons $\Lambda^{*}_{\bar{c}c}$ are also
predicted in Refs.~\cite{Wu:2010jy,Wang:2011rga}. $\Lambda^{*}(4544)$
decays to a vector meson and strange baryon or a $K^*$ meson and a
nucleon.  The VMD mechanism connect the vector with photon, so a strange
baryon is essential in the initial state to produce $\Lambda^*(4544)$.
Considered the $K^*$ meson can not be connected to a photon,
$\Lambda^{*}(4544)$ can not be produced in $\gamma p$ collision
through the mechanism in Fig 1(a). For the mechanism in Fig~\ref{Fig:
diagram}(b) the exchanged meson should be a strange meson for $\gamma
p$ collision. The zero decay width of $\Lambda^*(4544)$ in $K^*N$
channel exclude such production channel.  Similarly,
$\Lambda^{*}(4394)$ can not be produced through both mechanisms.
$\Lambda^{*}(4209)$ and $\Lambda^{*}(4368)$ can be produced by
mechanism (b).

\begin{figure}[h!]
\begin{flushright}
\includegraphics[width=0.85\textwidth]{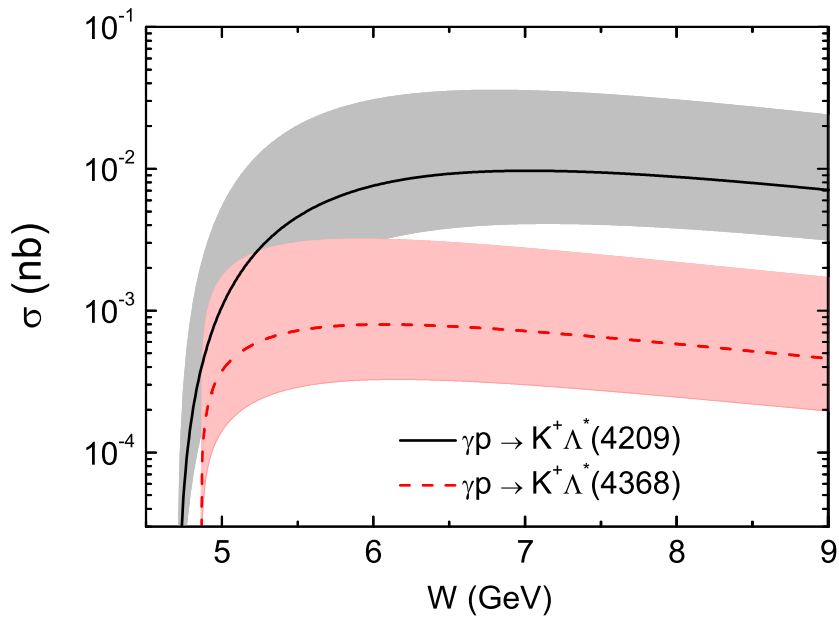}
\caption{(Color online) The total cross sections for the productions of
$\Lambda^{*}(4209)$ and $\Lambda^{*}(4368)$.  The bands sre for the uncertainties
with the variation of $\alpha$ from 1 to 2.
\label{Fig: Lcc} }
\end{flushright}
\end{figure}

The thresholds of $\Lambda^{*}(4209)$ and
$\Lambda^{*}(4368)$ productions are higher than the energy region of
JLab 12~GeV, but may be reached by the proposed EICs at HIAF and FAIR.
Here we present our results for the future possible experiments.  The
cross sections of the productions of $\Lambda^*_{c\bar{c}}$s are
plotted in Fig.~\ref{Fig: Lcc}. The total cross sections of $\Lambda^{*}(4209)$ and
$\Lambda^{*}(4368)$ productions are of
the order of $10^{-3}$~nb, which are smaller than these of hidden charm
nucleon resonances $N^*(4261)$ production because the exchanged mesons $K$ and
$K^*$ here are heavier than the exchanged meson $\pi$ in $N^*(4261)$ production .

\section{Summary}

The baryon resonances with hidden charm  $B^*_{c\bar{c}}$ with mass
above 4~GeV have been predicted in several approaches. The JLab
12~GeV and the proposed EICs will provide excellent experimental
condition to search for the hidden charm baryon resonances.
In this work, the possibility of observation of $D^*_{c\bar{c}}$resonances $via$ photoproduction are studied in the effective
Lagrangian approach with VMD mechanism.

A peak around 4.412 GeV can be found near the threshold for $J/\psi$ photoproduction .  The theoretical
values of the total cross sections of near threshold $J/\psi$
photoproduction through the hidden charmed nucleon resonance
$N^*(4412)$ is about 1~nb. If the precision of the experiment reach
the order of 0.1~nb,  it is promising to observe $N^*(4412)$. Such
requirement can be reached by JLab after 12~GeV upgrade. Besides, a
measurement of the differential cross section with large $t$ around
$W=4.4$ GeV is suggested.

The total cross section for the $N^*(4261)$ production is about one
order smaller than the total cross section for $N^*(4412)$ with large
background. The
thresholds for the production of hidden charm strange baryon
resonances $\Lambda^*_{c\bar{c}}$ are beyond the energy region of JLab 12~GeV and the total
cross sections are in the order $10^{-3}$~nb. Hence, it is difficult to
search for these states in JLab and the proposed EICs.

Based on the results above, the search for $N^*(4412)$ in the $\gamma p\to
J/\psi p$ channel is most promising. Our results will be helpful to the
experimental search for the hidden-charmed baryon resonances in the
coming experiment at JLab 12~GeV, such as SoLID, and the proposed
electron-ion colliders at FAIR and HIAF.

\ack
This project is partially supported by the Major State Basic Research
Development Program in China (No. 2014CB845405), the National Natural
Science Foundation of China (Grants No. 11275235, No. 11035006) and
the Chinese Academy of Sciences (the Knowledge Innovation Project
under Grant No. KJCX2-EW-N01).

\end{document}